\documentclass[aps, prl, twocolumn, nofootinbib, groupedaddress, amsfonts, amssymb, amsmath]{revtex4-1}
\usepackage{graphicx}
\usepackage{color}
\usepackage{amssymb}
\usepackage{amsmath}
\usepackage{tabularx}
\usepackage{bm}
\usepackage{hyperref}
\usepackage{ltxtable}
\usepackage{natbib} 
\usepackage{longtable}
\usepackage{ulem}
\usepackage[usenames,dvipsnames]{xcolor}
\graphicspath{{./fig/}}

\hypersetup{
colorlinks,
citecolor=blue,
filecolor=blue,
linkcolor=blue,
urlcolor=blue}
\def\apj{ApJ}
\def\mnras{MNRAS}
\def\memsai{Memorie della Societ\'a Astronomia Italiana}
\newcommand\aap{{Astrophys. J}}%
\newcommand\apjl{{Astrophys. J Lett.}}%
\def\avg{\bar}
\def\vec{\boldsymbol}
\newcommand{\del}{\mbox{\boldmath $\nabla$}}

\def\scrD{\mathcal{D}}
\def\scrR{\mathcal{R}}
\def\Rs{R$_{\odot}$}
\def\Pra{\mbox{\rm Pr}}
\def\Ta{\mbox{\rm Ta}}
\def\Pe{\mbox{\rm Pe}}
\def\Ra{\mbox{\rm Ra}}

\newcommand{\cp}{C_{\rm p}}

\newcommand{\Eq}[1]{Equation~(\ref{#1})}

\newcommand{\urms}{u_{\rm rms}}
\newcommand{\lcor}{l_{\rm corr}}
\newcommand{\Ro}{{\rm Ro}}

\newcommand{\mps}{m~s$^{-1}$}
\newcommand{\brac}[1]{\langle #1 \rangle}
\newcommand{\Fig}[1]{Figure~\ref{#1}}

\newcommand{\Tab}[1]{Table~\ref{#1}}

\begin{document}
\title{Consequences of high effective Prandtl number on solar differential rotation and convective velocity}
\author{Bidya Binay Karak$^{1,2}$}
\email{bidyakarak@gmail.com}
\author{Mark Miesch$^2$}
\email{miesch@ucar.edu}
\author{Yuto Bekki$^3$}
\email{bekki@eps.s.u-tokyo.ac.jp}
\affiliation{$^1$Indian Institute of Astrophysics, Koramangala, Bangalore 560034, India\\
$^2$High Altitude Observatory, National Center for Atmospheric Research, 3080 Center Green Dr., Boulder, CO 80301, USA\\
Department of Earth and Planetary Science, The University of Tokyo, 7-3-1 Hongo, Bunkyo-ku, Tokyo 113-0033, Japan}
\date{\today}

\begin{abstract}
Observations suggest that the large-scale convective velocities obtained by solar convection simulations might be over-estimated (convective conundrum). 
One plausible solution to this could be the small-scale dynamo which cannot be fully resolved by global simulations. 
The small-scale Lorentz force suppresses the convective motions and also the turbulent mixing of entropy between upflows and downflows, leading to a large effective Prandtl number ($\Pr$). 
We explore this idea in three-dimensional global rotating convection simulations at different thermal conductivity ($\kappa$), i.e., at different $\Pr$.
In agreement with previous non-rotating simulations, the convective velocity is reduced with the increase of $\Pr$ as long as the thermal conductive flux is negligible. 
A subadiabatic layer is formed near the base of the convection zone due to continuous deposition of low entropy plumes in low-$\kappa$ simulations. 
The most interesting result of our low-$\kappa$ simulations is that the convective motions are accompanied by a change in the convection structure that is increasingly influenced by small-scale plumes. 
These plumes tend to transport angular momentum radially inward and thus establish an anti-solar differential rotation, in striking contrast to the solar rotation profile. 
If such low diffusive plumes, driven by the radiative-surface cooling, are present in the Sun, then our results cast doubt on the idea that a high effective $\Pr$ may be a viable solution to the solar convective conundrum. 
Our study also emphasizes that any resolution of the conundrum that relies on the downward plumes must take into account angular momentum transport as well as heat transport.
\end{abstract}


\maketitle
\section{Introduction}\label{sec:intro}
The outer layer of cool stars like the Sun is known to be convectively unstable whereby
the energy from the interior to the surface is transported through convection. 
This, along with the global rotation of the star, drives differential rotation, meridional circulation, and the dynamo action responsible for generating magnetic activity \citep{Miesch05,Kar14a}.

Although we have a tremendous success in modeling solar convection at granular scales
through local Cartesian simulations \cite{NSA09}, we face unavoidable challenges in simulating
the global convection encompassing the full convection zone (CZ). It has been realized that simulations
produce substantially higher convective velocities at large scales (small horizontal wave numbers) 
than that inferred from photospheric and helioseismic observations \citep{Hanasoge12,Lord,Greer}.
In fact, numerical models, which reproduce the solar surface granulations remarkably well,
when extended to a larger extent, tend to produce larger power at low spectral wavenumber
in comparison to photospheric observations \citep{Lord}. This discrepancy between the observations
and models is commonly known as the convective conundrum; see Section~1 of \citet{Omara} for 
a detailed discussion.

A possible solution to the convective conundrum could be that the solar
convection is driven by a non-local process. As argued by \citet{Spr97}, the convection is
driven by the efficient cooling at the surface. The low entropy plumes generated near the surface
penetrate the deeper CZ, in a process known as {\it entropy rain} \citep{Br16,CR16}.
The nonlocal heat transport of these downward plumes maintains a subadiabatic layer where the enthalpy flux is
still positive \citep{Br16}. The existence of such a subadiabatic layer has been confirmed in a local Cartesian simulations with
a smoothly varying heat conduction at the lower CZ based on either a Kramers-like opacity 
which is a function of temperature and density or a static profile of a similar shape \citep{Kap17}; 
also see \citet{Kap18b} for the extension of this study in the spherical geometry with dynamo.

Another possible solution of the convective conundrum, as proposed by \citet{HRY15}, could be the
suppression of the convective velocity due to the Lorentz force of the dynamo-generated magnetic field.
In many global MHD simulations, it has been seen that the magnetic field reduces the convective velocity
through the suppression of shear and thus effectively increasing the
turbulent viscosity ($\nu_{\rm eff}$) \citep{FF14,Kar15,Kap17a}.
The effect of this has been realized in the transition to the solar-like differential rotation 
from the anti-solar profile \citep{FF14,Kar15}.
Solar-like differential rotation is characterized by 
faster equator and slower poles (equatorward $\nabla \Omega$). The opposite scenario 
where the equator rotates slower than the poles (poleward $\nabla \Omega$), is referred to as the 
anti-solar (AS) differential rotation.
In simulations cited above, the magnetic field is produced only from the large-scale dynamo. However,
the small-scale dynamo also produces an immense amount of magnetic flux in the whole CZ and thus the effect could be profound \citep{Rem14,KB16,Kap17a,Kap18a}.
Recently, Hotta et al.\ \citep{HRY15,HRY16} carried out a set of global convection
simulations with and without rotation and found about $50\%$ reduction in convective velocities due to the small-scale magnetic field in their highest resolution simulation that they could achieve.
Another effect of magnetism is that it reduces the turbulent mixing of entropy fluctuations in downflow plumes.
This decreases the effective thermal diffusivity ($\kappa_{\rm eff}$). 
Therefore, the magnetism effectively increases the turbulent Prandtl number $\Pr_{\rm eff} (= \nu_{\rm eff}/\kappa_{\rm eff}$) if the convection is essentially magnetized.

Since any global simulations at present cannot fully capture the small-scale dynamo action as realized in the real Sun, the effects are often modeled as an enhanced effective Prandtl number $\Pr_{\rm eff}$ to investigate the propoerties of the magnetized stellar convection.
At this end, \citet{Omara} carried out a set of convection simulations at varying $\Pr$ and they showed that the convective velocity decreases with the increase of $\Pr_{\rm eff}$.
In local Cartesian simulations, \citet{Bekki} also find a similar result of decreasing convection velocity
with the increase of $\Pr$. They further demonstrated that 
a subadiabatic layer near the base of the CZ is formed by continuous deposition of low-entropy downward plumes.
The depth of this layer increases with the decrease of horizontal thermal diffusivity.

If solar convection is largely driven by entropy rain, caused by the radiative surface cooling
and/or by the large effective $\Pr$ caused by the magnetic field, then this is certainly an encouraging thought
because both of these effects act to reduce the convective velocity. We may hope that in a realistic scenario these effects might
solve the solar convective conundrum. However, the problem is more subtle. These downward plumes can also transport angular momentum inward
and thus produce a radially decreasing rotation rate i.e., an anti-solar differential rotation. 
As the simulations mentioned above \citep{HRY15,HRY16,Omara,CR16,Bekki,Kap17} were all performed in local geometry without rotation,
this effect could not be explored.

In this study, we explore the effect of downward plumes in forming the subadiabatic layer and particularly 
in transporting angular momentum in rotating convection.
We perform a few sets of rotating and non-rotating global
convection simulations in spherical geometry at different $\Pr$. We find that at larger $\Pr$ the convective velocity is suppressed and a subadiabatic layer is formed near the base of the CZ due to continuous deposition of low entropy plumes. However, our results are quantitatively different than the previous non-rotating (local) Cartesian simulations \citep{Omara,Bekki}. The most interesting and unforeseen result of our simulations is that the inward transport of angular momentum by plumes leads to an anti-solar differential rotation in high-$\Pr$ regime, despite the stronger rotational influence as quantified by the lower Rossby number.

\section{Numerical Model}\label{sec:model}
In our study we solve the hydrodynamic (non-magnetic) equations in rotating spherical geometry under the anelastic approximation
using the pseudo-spectral code, Rayleigh; see \citet{FH16} for details.
With this approximation, thermodynamic variables are linearized about a spherically symmetric, 
time-independent 
reference state with density $\avg{\rho}$, pressure $\avg{P}$, temperature $\avg{T}$, 
and specific entropy $\avg{S}$ and fluctuations about this state are represented by $\rho$, $P$, $T$, and $S$, respectively.  
The reference state is assumed to be in an adiabatically-stratified hydrostatic equilibrium; see Appendix C of \citet{FH16} for more details.

With this approximation, the continuity equation reduces to 
\begin{equation}  
  \label{eq:continuity}
  \del \cdot(\avg{\rho}\vec{v}) = 0,
\end{equation}
where $\vec{v}$ is the velocity field.  
The momentum equation is given by
\begin{equation}  
  \label{eq:momentum}
\begin{split}
  \avg{\rho} \left[ \frac{D\vec{v}}{Dt} + 2{\bm \Omega_0}\times{\bm v}\right]
  =
  -\avg{\rho}\del\frac{P}{\avg{\rho}} - \frac{\avg{\rho}S}{c_p} \vec{g} - \del \cdot \scrD.
\end{split}
\end{equation}
Here $\vec{g}$ is the gravitational acceleration, $c_p$ is the specific heat at constant pressure, 
$\bm\Omega_0=(\cos\theta,-\sin\theta,0)\Omega_0$ is the angular velocity vector, and
the viscous stress tensor $\scrD$ 
is given by
\begin{equation}  
\label{eq:stress}
  \scrD_{ij} = -2 \avg{\rho} \nu \left[e_{ij}
    - \frac{1}{3}(\del \cdot \vec{v})\delta_{ij} \right],
\end{equation}
where $e{_{ij}}$ is the strain rate tensor, $\nu$ is the kinematic viscosity, and $\delta_{ij}$ 
is the Kronecker delta. 
The energy equation is given by
\begin{eqnarray}
  \label{eq:entropytwo}
  \begin{split}
  \avg{\rho}\avg{T}\frac{DS}{Dt} = 
  \del \cdot [\kappa \avg{\rho} \avg{T} \del S] 
                    + 2 \avg{\rho}\nu \left[e_{ij}e_{ij} - \frac{1}{3}(\del \cdot  \vec{v})^2\right]\\
+ Q_\mathrm{heat} + Q_\mathrm{cool},
  \end{split}
\end{eqnarray}
where the thermal diffusivity is denoted by $\kappa$, and $Q_\mathrm{heat}$ and $Q_\mathrm{cool}$ are the internal heating and cooling.
Following previous studies \citep{FH16,Omara}, we use 
a functional form of $Q_\mathrm{heat}$ that depends only on the background pressure profile such that
\begin{equation} 
\label{eq:heating}
Q_\mathrm{heat}(r) = \alpha \left[\avg{P}(r)-\avg{P}(r_{o})\right].
\end{equation}
The normalization constant $\alpha$ is chosen so that
\begin{equation}
L_\odot = 4\pi \int_{r_i}^{r_o} r^2 Q_\mathrm{heat}(r) dr, 
\end{equation}
where $L_\odot$ is the stellar luminosity, $r_i$ and $r_o$ are the inner and outer radii of the computation domain. 
The radial dependence of the net radiative heat flux
(see, for example, Figure 6 of \citet{FH16}) 
is then defined as
\begin{equation} 
\label{eq:heatflux}
F_\mathrm{r}(r) = \frac{L_\odot}{4\pi r^2} - \frac{1}{r^2}\int_{r_i}^{r} {r^{\prime}}^2 Q_\mathrm{heat}(r^{\prime}) dr^{\prime}.
\end{equation}

For $Q_\mathrm{cool}$, we use the same form as given in \citet{HRY14} and \citet{Bekki} i.e.,
\begin{equation} 
\label{eq:cooling}
Q_\mathrm{cool}(r) = -\frac{1}{r^2} \frac{\partial}{\partial{r}}(r^2 F_\mathrm{s}),
\end{equation}
where 
\begin{equation}
\label{eq:coolflux}
F_\mathrm{s} = \frac{L_\odot} {4\pi r^2} \exp{\left[-\left(\frac{r-r_o}{H_\mathrm{p}}\right)^2\right]},
\end{equation}
with $H_\mathrm{p}$ being the pressure scale height.

Finally, the linearized equation of state is given by
\begin{equation} 
  \frac{\rho}{\avg{\rho}} = \frac{P}{\avg{P}} - \frac{T}{\avg{T}}
    =  \frac{P}{\gamma \avg{P}} - \frac{S}{c_p},
\end{equation}
which is obtained by assuming the ideal gas law
$  \avg{P} = \scrR \avg{\rho} \avg{T}$,
where $\gamma$ is the adiabatic index of the gas
and $\scrR$ is the gas constant.

Our model is constructed using an adiabatic (constant $\overline{S}$), polytropic background state, which is specified 
with seven inputs, namely $r_i =  0.7$\Rs, $r_o = R_\odot = 6.96\times10^{10}$~cm, 
the polytropic index $n_p = 1.5$,
the number of density scale heights within the domain $N_\rho = 3$ 
(see the green curve in Figure 1a of \citet{FH16}),
 the mass below the inner boundary 
$M_i = 1.989\times10^{33}$~gm, the density at the inner boundary $\rho_i = 0.1805$~g~cm$^{-3}$,
and $c_p = 3.5\times10^8$~erg~g$^{-1}$~K$^{-1}$.

In our model at both boundaries, we impose impenetrable and stress-free conditions such that 
\begin{equation}
v_r =\frac{\partial}{\partial r}(\frac{v_\theta}{r})=\frac{\partial}{\partial r}(\frac{v_\phi}{r})=0|_{r = r_i,r_o}.
\end{equation}
For entropy we have the following conditions: 
\begin{equation}
\label{eq:sboundary}
\left.\frac{\partial S}{\partial r} \right|_{r=r_i}= 0,~~S(r_o) = 0.
\end{equation}

Thus there is no diffusive entropy flux across the bottom boundary. 
However, the convection is driven by depositing heat in the domain through $Q_\mathrm{heat}$
(\Eq{eq:heating}) which decreases to zero at the outer boundary and by cooling the upper part through $Q_\mathrm{cool}$ (\Eq{eq:cooling}).

\section{Results}\label{sec:results}
We perform three sets of simulations, namely, at the solar rotation rate (R1), five times solar rotation (R5) and without rotation (R0).
In each set, we vary the Prandtl number $\Pr$ from 1 to 20 by decreasing the thermal diffusivity $\kappa$ alone.
Each run is labeled with RxPy which represent a simulation at x times solar rotation rate and at $\Pr=$ y; see \Tab{table1}.
All simulations are run for several thermal diffusion times and we consider the results only
after they have reached the thermally relaxed state.
During this state, the flow remains statistically stationary.

\subsection{Formation of Subadiabatic Layer}
\label{sec:sub}
\begin{figure}
\includegraphics[scale=0.4]{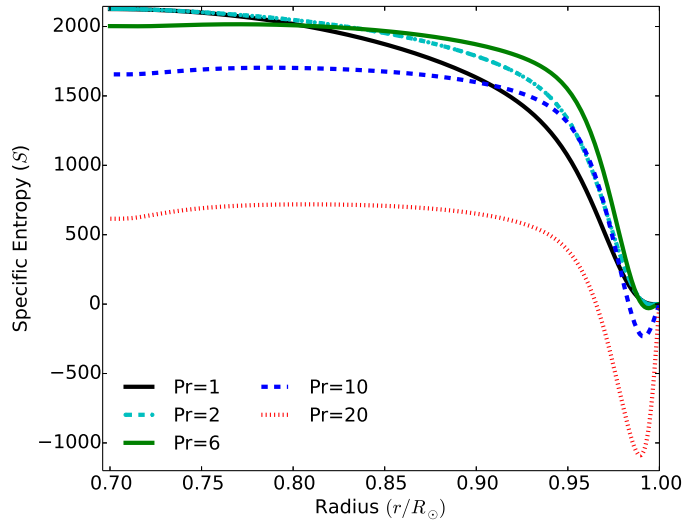}
\caption{Radial dependence of the specific entropy $\brac{S}_{\theta,\phi,t}$ in unit of erg~g$^{-1}$K$^{-1}$
obtained from Runs~R1P1 (black), R1P2 (cyan), R1P6 (green), R1P10 (blue), and R1P20 (red).
}
\label{fig:Svsr}
\end{figure}

\Fig{fig:Svsr} shows the specific entropy $\brac{S}_{\theta,\phi,t}$ averaged over time and the horizontal dimensions 
as a function of radius, from the R1 set of simulations (R1P1--R1P20).
We observe that in most of the CZ, particularly for the larger $\Pr$ runs, the entropy stratification is close to adiabatic
and at the surface it goes to zero due to the imposed boundary condition.
The cooling layer near the surface tries to reduce the entropy just below the surface and 
this causes a narrow subadiabatic layer (with $dS/dr > 0$) there.
This effect is more pronounced at smaller $\kappa$ (larger $\Pr$) because the thermal diffusion is insufficient to cause the local cooling to spread.  The resulting buoyant acceleration then establishes the convection.
Similar behavior is also seen in the previous local non-rotating simulations of \citet{Bekki},
although they do not have fixed entropy boundary condition at the top.

\begin{table*}
\begin{ruledtabular}
\caption{Simulation summary.
In all simulations, $\nu$ is fixed at $8\times10^{12}$~cm$^2$~s$^{-1}$.
Thus in each set of simulations, the Prandtl number 
$\mathrm{Pr} = \nu/\kappa$ is varied by changing $\kappa$. 
The Taylor number, $\Ta = (2 \Omega_0 d^2/\nu)^2$ with $d = 0.3$\Rs, in the R1 set of simulations is $8\times10^4$,
while in R5 set it is $2\times10^6$.
Other parameters are defined as follows:
the Rayleigh number
$\mathrm{Ra} = \frac{G M_{\odot} d^3}{R_\odot^2 \nu \kappa \cp}\Delta{S}$,
where $\Delta S = \brac{S}_{\theta,\phi}|_\mathrm{max} - \brac{S}_{\theta,\phi}|_\mathrm{min}$ 
(in unit of erg~g$^{-1}$K$^{-1}$)
is the entropy contrast across the layer.  
The rms value of the convective velocity 
$\urms = \brac{ \sqrt{ (v_r-\brac{v_r}_\phi)^2+(v_\theta-\brac{v_\theta}_\phi)^2+(v_\phi-\brac{v_\phi}_\phi)^2} }_{\{r,\theta,\phi,t\}}$ 
in \mps,
the convective Rossby number
$\Ro_c = \sqrt{\frac{\Ra}{\Pra \Ta}}$, 
Rossby number
$\Ro = \urms /(2\Omega_0 \lcor)$
with $\lcor = d/2\pi $ is an estimate of the largest eddies,
$\Pe = \urms \lcor / \kappa$,
$L_\mathrm{e}^\mathrm{p}$ is peak of the enthalpy flux, 
 $d_{\rm{sa}}$ is the spherically-averaged width of the subadiabatic layer (with $\delta < 0$) in Mm,
formed in the lower CZ in the statistically stationary state,
$f_d$ is the filling factor of the downflow,
$\overline{\rm RS}_\mathrm{r}, \overline{\rm RS}_\theta$ respectively show the strength of the radial and latitudinal 
Reynolds stresses averaged over the northern hemisphere in units of $10^{15}$~g~s$^{-2}$,
$\Delta_\Omega^{\mathrm (r)} = \Omega(R_\odot,90^\circ) -\Omega(0.71R_\odot,90^\circ)$, 
and $\Delta_\Omega^{\mathrm (\theta)} = 0.5 \left[2 \Omega(R_\odot,90^\circ)- \Omega(R_\odot,30^\circ) - \Omega(R_\odot,120^\circ)\right]$.
}
\begin{tabular}{lccrclcccccccc}
Run  &$\Pr$ & N$_r$, N$_\theta$, N$_\phi$ & $\Delta S$ & $\Ra[10^4]$ &$\Ro_c$& $\urms$ (m/s) & $\Ro$ & Pe & $L_\mathrm{e}^\mathrm{p}$ & d$_{\rm{sa}}$ (Mm) &$f_d$&$\overline{\rm RS}_\mathrm{r}, \overline{\rm RS}_\theta$ & {$\Delta_\Omega^{\mathrm (r)}$}, {$\Delta_\Omega^{\mathrm (\theta)}$} \\ \hline
R1P1 & 1 & 128,192,384  & 2282 & $2.5$ &0.56&77.7 & 0.45 & 20.3 & 0.65 & ~0.0 &0.44& $~~1.48,6.8$& ~~14.8,   ~41.5\\
R1P2 & 2 & 128,256,512  & 2319 & $5.2$ &0.57&91.7 & 0.53 & 47.9 & 0.94 & ~0.0 &0.42& $-1.91,6.8$ & ~~~8.8,   ~21.9\\
R1P6 & 6 & 128,512,1024 & 2477 & $17$ &0.59&91.2 & 0.53 & 143  & 1.02 & 45.3 &0.42& $-6.03,4.7$ &$-15.8$,$-2.0$\\
R1P10& 10& 128,1024,2048& 2499 & $28$ &0.59&85.9 & 0.50 & 224  & 1.01 & 60.9 &0.42& $-6.41,3.6$ &$-20.7$,$-3.6$\\
R1P20& 20& 200,2048,4096& 2389 & $53$ &0.58&77.7 & 0.45 & 406  & 0.98 & 77.6 &0.42& $-6.25,2.7$ &$-24.4$,$-2.1$\\
\hline
R5P1 & 1 & 128,192,384  &4086& $4.6$ & 0.15 &10.4 & 0.012 &~2.7& 0.03 &~0.0 &0.37& $~~0.04,0.1$ & ~1.2, ~2.2 \\
R5P2 & 2 & 128,192,384  &6380& $14$ & 0.19 &40.7 & 0.047 &21.2& 0.41 &~0.0 &0.49& $~~1.73,2.3$ & 19.9, 44.2 \\
R5P6 & 6 & 128,512,1024 &5271& $35$ & 0.17 &59.9 & 0.069 &93.8& 0.88 &~0.0 &0.47& $~~2.25,3.6$ & 22.6, 57.3 \\
R5P20& 20& 200,2048,4096&4412& $99$ & 0.16 &54.2 & 0.062 &283 & 0.92 &20.4 &0.46& $ -0.75,1.4$ & ~7.3, 28.3 \\
\hline
R0P1 & 1 & 128,192,384  &1886& $2.1$ &$\infty$&140.8 &$\infty$&~36.8 & 1.17 &90.4&0.31&\ldots&\ldots\\
R0P2 & 2 & 128,192,384  &2196& $4.9$ &$\infty$&130.6 &$\infty$&~68.2 & 1.19 &90.4&0.33&\ldots&\ldots\\
R0P6 & 6 & 128,512,1024 &1972& $13$ &$\infty$&111.0 &$\infty$&174.1 & 1.02 &92.9&0.35&\ldots&\ldots\\
R0P20&20 & 200,2048,4096&2100& $48$ &$\infty$&~80.5 &$\infty$&420.0 & 1.00 &95.4&0.35&\ldots&\ldots\\
\end{tabular}
\label{table1}
\end{ruledtabular}
\end{table*}


\begin{figure*}
\includegraphics[scale=0.4]{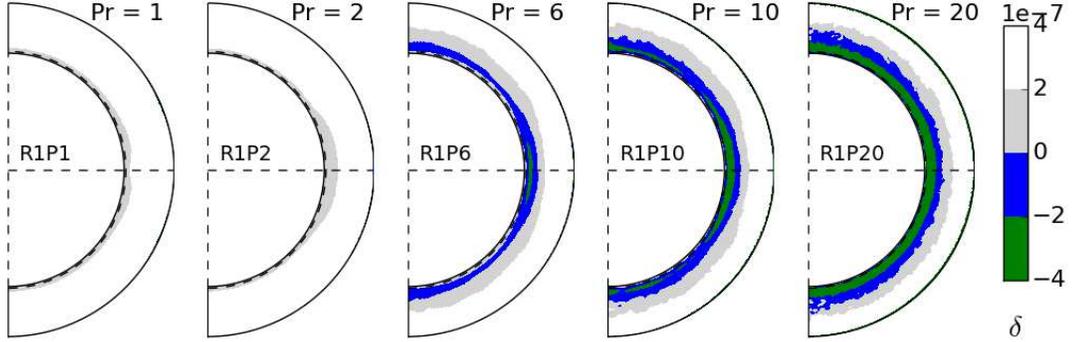}
\caption{
Superadiabaticity $\delta = -(H_{\mathrm p}/c_{\mathrm p}) \frac{d}{dr}\brac{S}_{\theta,\phi,t}$ 
obtained by averaging over azimuth and time from Runs~R1P1--R1P20.
Note the blue and green colors show the subadiabatic layer ($\delta < 0$) developed in the statistically stationary convection.
}
\label{fig:dsdr}
\end{figure*}

\begin{figure*}
\includegraphics[scale=0.4]{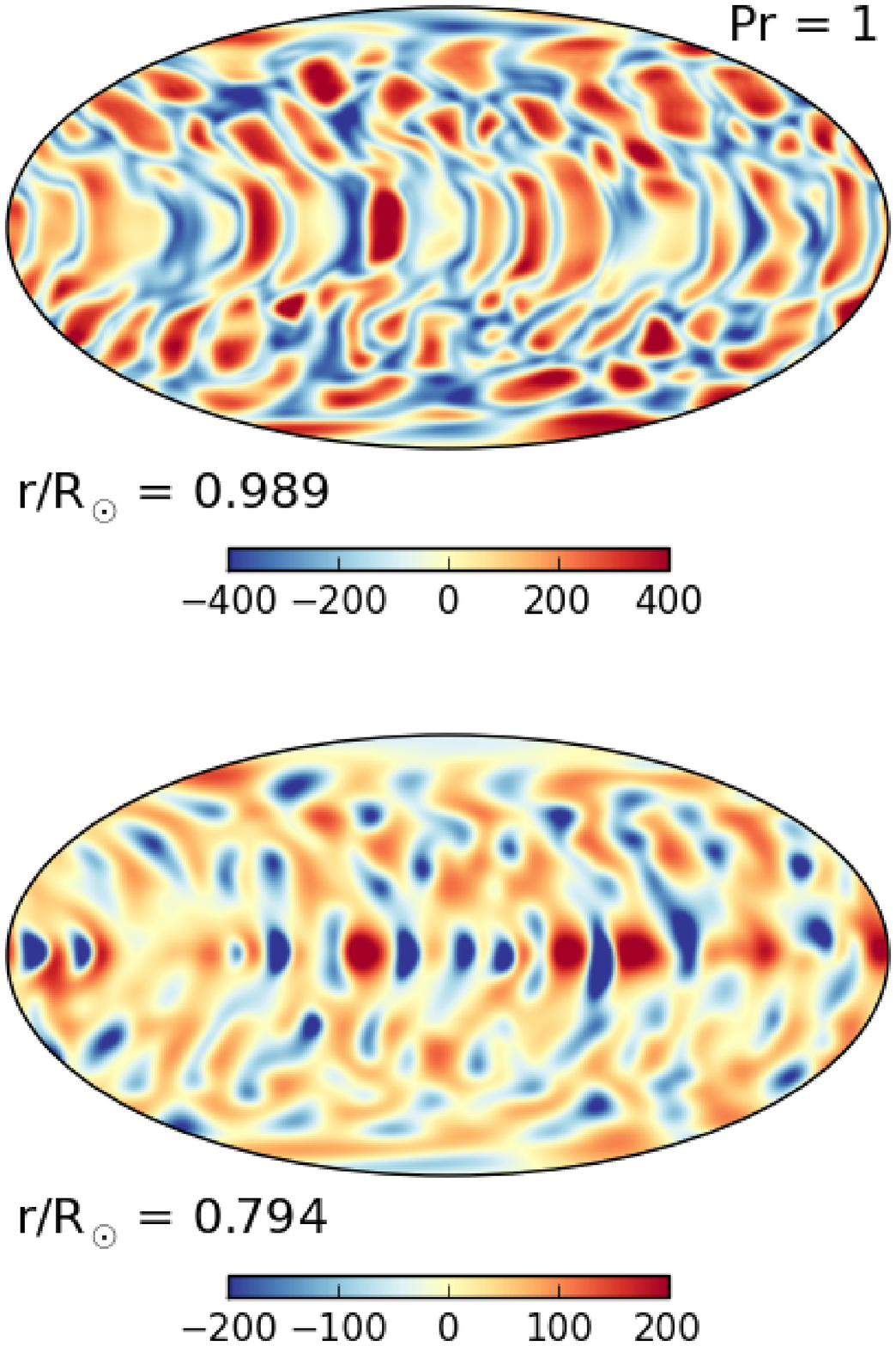}
\includegraphics[scale=0.4]{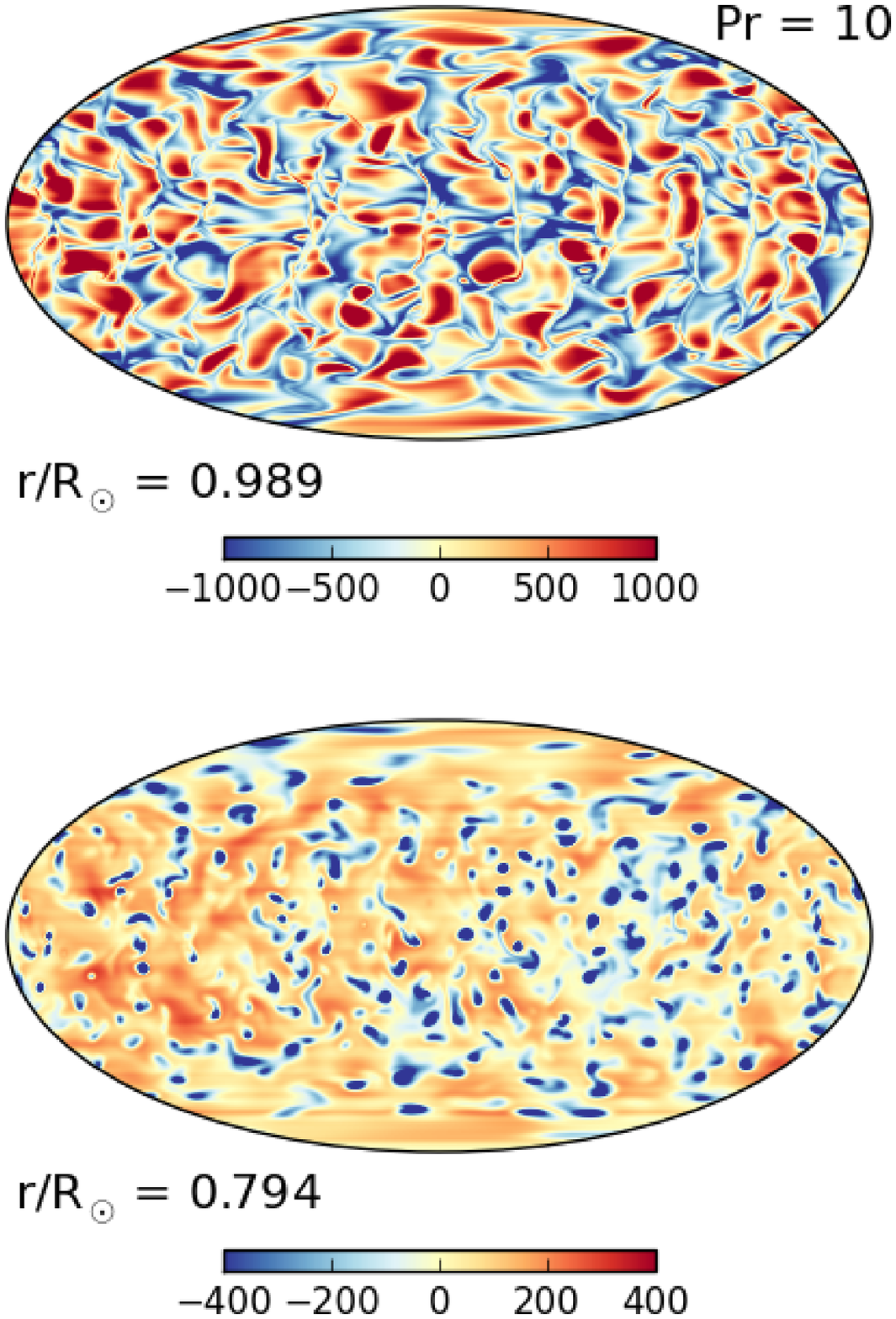}
\caption{Entropy fluctuation $S - \brac{S}_\phi$ in erg~g$^{-1}$K$^{-1}$
at two different radial layers 
(top panels: near the upper boundary; bottom: near the base of the CZ) 
from Run~R1P1 (left panels), and Run~R1P10 (right panels).}
\label{fig:Spert}
\end{figure*}

As $\Pr$ is increased, we start getting a 
positive 
mean entropy gradient in the lower part of the domain
and thus a subadiabatic layer in which $dS/dr > 0$ is formed.
In \Fig{fig:dsdr}, we show the subadiabaticity $\delta$ which is defined as 
\begin{equation}  
  \label{eq:super}
  \delta = \Delta - \Delta_{\mathrm{ad}} = -\frac{H_{\mathrm p}}{c_{\mathrm p}} \frac{d}{dr}\brac{S}_{\theta,\phi,t},
\end{equation}
where $\Delta = d \ln T/d \ln P$ is the logarithmic temperature gradient and $\Delta_\mathrm{ad}$ is the adiabatic one.
This subadiabatic
layer formed in our simulations is a consequence of the accumulation of low entropy plumes. 
As demonstrated by \citet{Bekki} \citep[also see][]{Kap17,CR16},
when $\kappa$ is small, the horizontal thermal diffusion of low entropy plumes is reduced and thus they can travel much deeper into the CZ before mixing with the surrounding fluid.  This is clearly seen in \Fig{fig:Spert} for the entropy perturbation. For $\Pr=10$,
the plumes reach all the way to the bottom of the CZ without losing their thermal content 
and the convection in this case is plume-like, while in the case of $\Pr=1$, 
the plumes are wider and less prominent, appearing together with columnar downflow lanes (banana cells).
Thus at larger $\Pr$, the downflow plumes accumulate in the lower CZ, forming the subadiabatic stratification.
In a statistically stationary state, the accumulation of low entropy plumes is compensated 
by vertical advection and diffusion.

\begin{figure}
\includegraphics[scale=0.4]{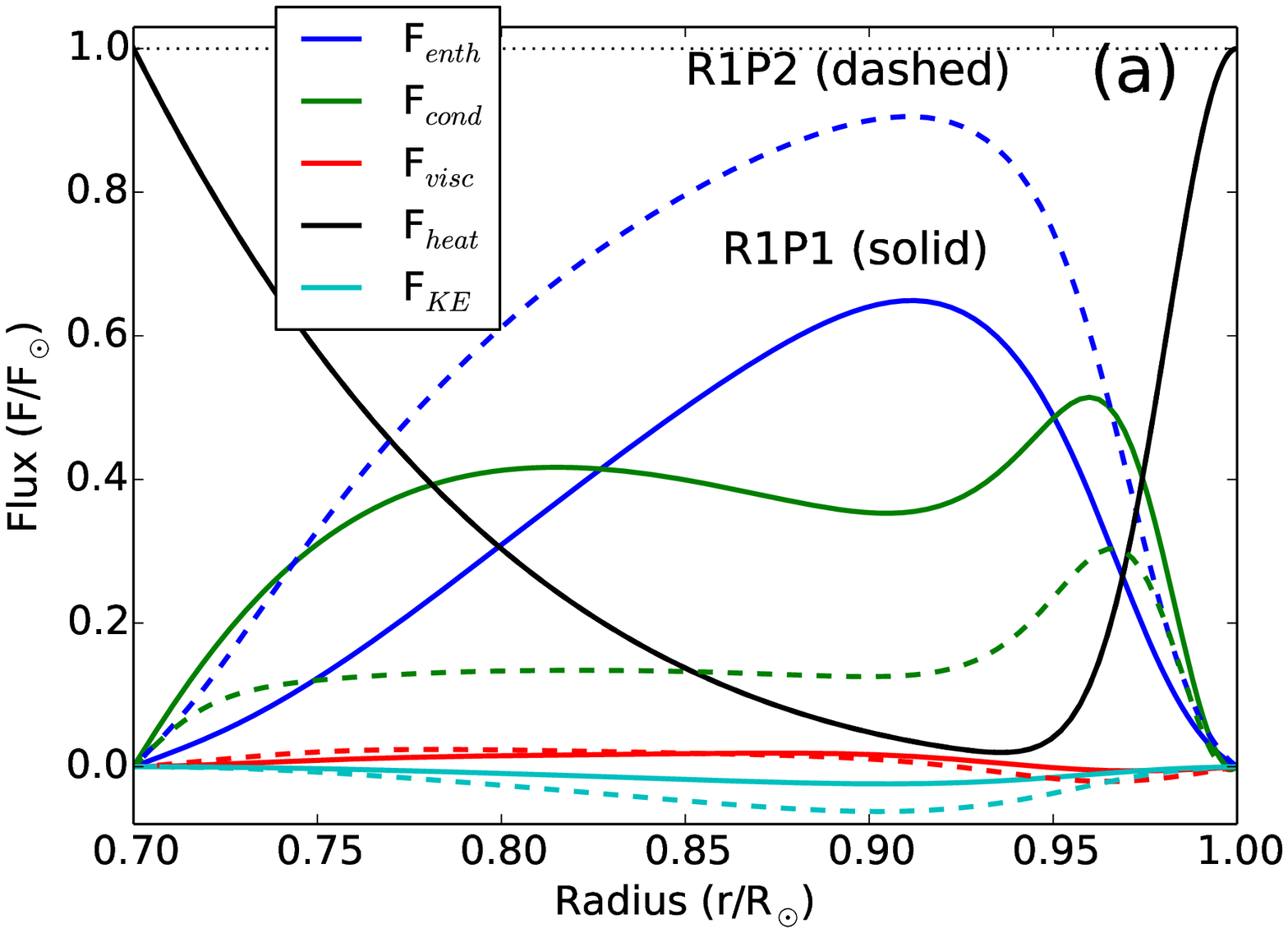}
\includegraphics[scale=0.4]{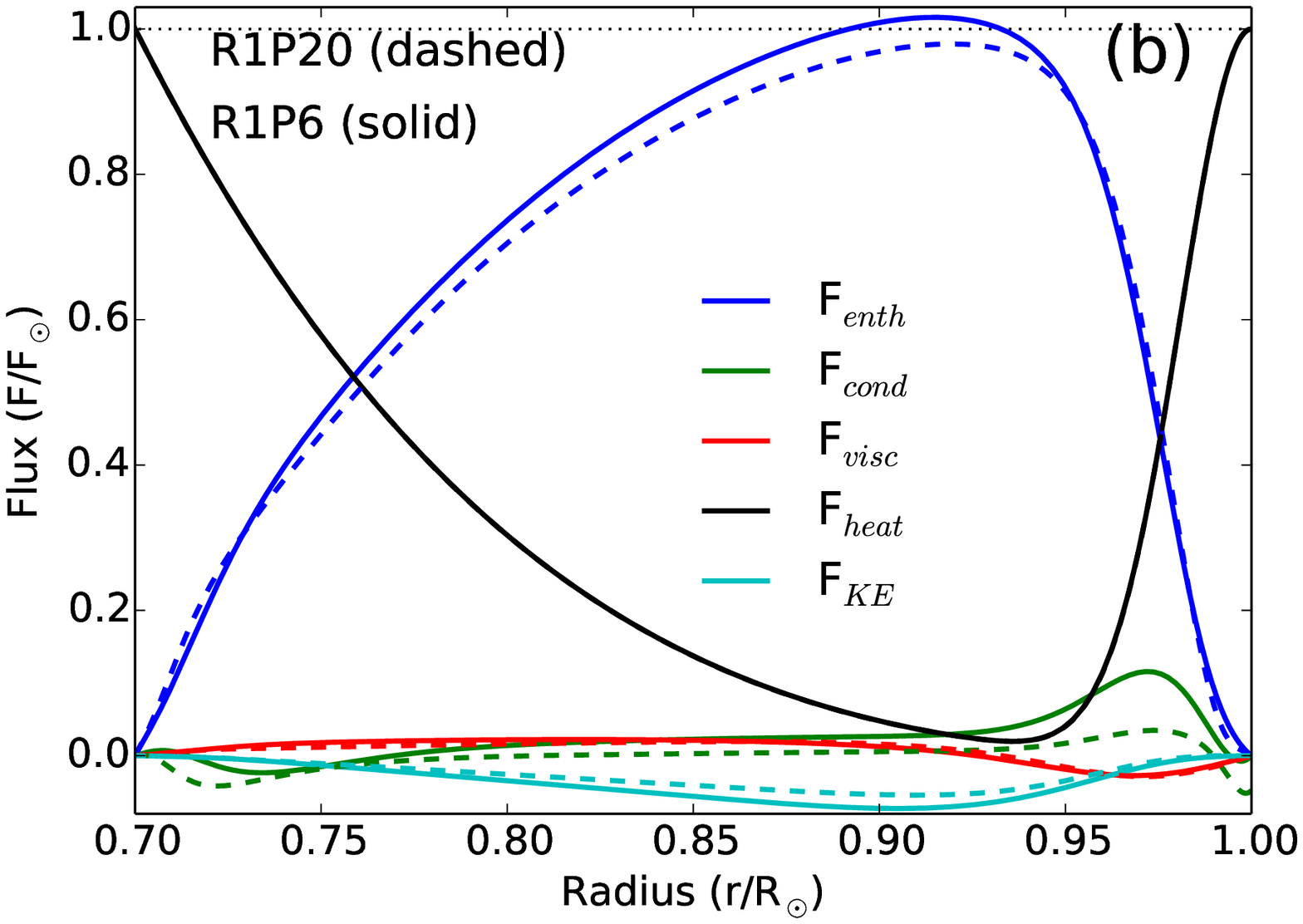}
\caption{Radial variations of enthalpy flux (blue), kinetic energy flux (cyan), conductive flux (green), 
viscous flux (red), and net heating and cooling flux $F_\mathrm{r}+F_\mathrm{s}$ (black) 
which includes the heating (decreasing monotonically to zero at the surface)
and cooling (significant only near the surface).
All fluxes are normalized by $F_\odot = L_\odot / 4 \pi r^2$. See \citet{FH16} for the definitions of these fluxes. 
(a) Obtained from Runs~R1P1 (solid) and R1P2 (dashed). (b) From Runs~R1P6 (solid) and R1P20 (dashed).}
\label{fig:fluxes}
\end{figure}

Interestingly, the enthalpy flux ($F_{\rm enth}$) is positive even near the base of the CZ where the subadiabatic
layer is formed; see \Fig{fig:fluxes}.
Thus in this subadiabatic layer the thermal energy is transported upward, which clearly manifests the nonlocal convective heat transport, possibly in a form of the non-gradient ({\sl Deardorff}) flux as discussed in \citet{Br16} and \citet{Kap17}.
This layer is distinct from
the overshoot region where the plumes are deaccelerated due to buoyancy and energy is transported downward.

\begin{figure}
\includegraphics[scale=0.45]{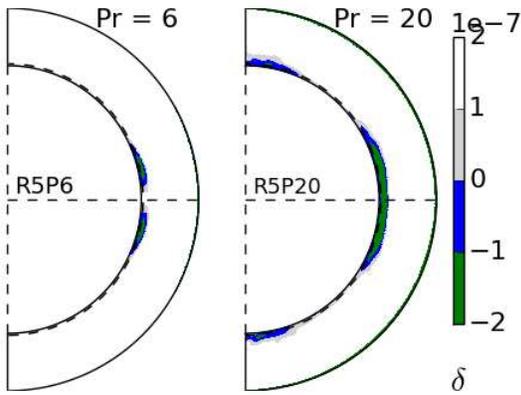}
\caption{
Similar to \Fig{fig:dsdr} but obtained from Runs~R5P6 (left) and R5P20 (right).
}
\label{fig:dsdrR5}
\end{figure}

We note that in non-rotating simulations, \citet{Bekki} detected the subadiabatic layer even at
$\Pr$ 1 and 2, for which we have not found any such layer. Although their simulations are performed
in local geometry and with a different surface boundary condition, we suspect rotation
to be the major cause of producing distinct results.
To explore this, we analyze the non-rotating simulations.
We find that for all values of $\Pr$, starting from 1, we get a subadiabatic layer; see Runs~R0P1--R0P20 in \Tab{table1}.
Therefore, the rotation which is not included in previous simulations of \citet{Bekki}
is the cause of the difference between their results and ours.
The Coriolis force inhibits the downward propagation of the plumes, 
suppressing the formation of the subadiabatic layer.
Due to the same reason, the R5 series of simulations ($\Omega_0 = 5 \Omega_\odot$)
show a thin subadiabatic layer only at $\Pr$ 6 and 20; see \Fig{fig:dsdrR5}.

Plumes at different latitudes feel a different effect of the Coriolis force.
Near poles, plumes approach the subadiabatic layer almost vertically,
while in low latitudes they travel at an angle or are subsumed into banana cells. 
Hence, we expect the extent of the subadiabatic layer
to decrease as we move away from the poles. Interestingly, in the R1 set of simulations, 
we do not find any significant latitudinal variation of the subadiabatic layer (\Fig{fig:dsdr}).
However, in the R5 set of simulations, we do see some latitudinal variation,
although not monotonic; see \Fig{fig:dsdrR5}.
The reason is that the dominant structure of rapidly rotating convection transitions from banana cells at low latitudes to plumes at high latitudes. 
As in previous rotating global convection simulations at moderate $\Ra$, 
convection is least efficient in the mid-latitude transition region \citep{Miesch05,FH16b}. 
Here this lower efficiency is manifested as the absence of a subadiabatic layer at mid-latitudes.

A similar phenomenon has been seen previously in the local f-plane simulations of \citet{BCT02} 
and the global convection simulations of \citet{Miesch00}.  Both found that the convective overshoot region was wider near the equator and poles and thinner at mid-latitudes.  However, their simulations included a radiative zone with a highly subadiabatic stratification and the overshoot region was characterized by a negative value of the convective enthalpy flux.   By contrast, the subadiabatic stratification in our simulations is relatively weak and the enthalpy flux is everywhere positive (radially outward).

\begin{figure}
\includegraphics[scale=0.45]{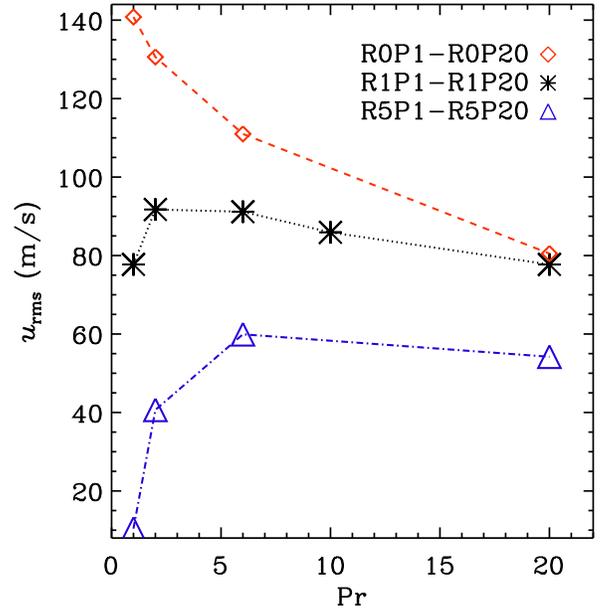}
\caption{$\Pr$ dependence of $\urms$. 
Asterisks/black: simulations at solar rotation (R1P1--R1P20).
Triangles/blue: five times solar rotation case (R5P1--R5P20).
Diamonds/red: non-rotating case (R0P1--R0P20)
}
\label{fig:urms}
\end{figure}

\begin{figure}
\includegraphics[scale=0.4]{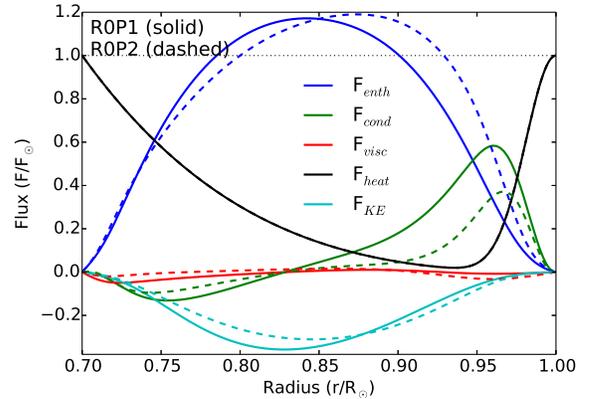}
\caption{Similar to \Fig{fig:fluxes}(a) but obtained from Runs~R0P1 (solid lines) and R0P2 (dashed).
}
\label{fig:fluxesR0}
\end{figure}

\subsection{Convective Velocity Amplitudes}
\label{sec:urms}
The root mean square of the convective velocity $\urms$,
averaged over the whole computation domain is shown in \Fig{fig:urms}.
In non-rotating simulations,  $\urms$ decreases with the increase of $\Pr$,
in agreement with previous findings \citep{Omara,Bekki}.
In rotating simulations, we find that $\urms$ increases first and
then decreases at larger $\Pr$ but at a slower rate than the non-rotating case.

The convective velocity suppression in the non-rotating case can be understood
following the discussion presented in \citet{Omara} (Sec.\ 5.2--5.4).
Based on the flux balance presented in \Fig{fig:fluxesR0} for Runs~R0P1 and R0P2,
we can write, the total flux = $L_\odot / 4 \pi r^2$  $\approx F_{\rm enth} + F_{\rm cond} + F_{\rm heat}$.
We notice that the enthalpy flux ($F_{\rm enth}$) is far more dominating over other fluxes in the mid CZ.
The heat flux $F_{\rm heat}$ which also includes the cooling near surface 
does not change in different runs and thus we shall omit it from the discussion here.
The kinetic energy flux $F_{\rm KE}$ is not negligible in this set of simulations, but
this does not change apperciably with the increase of $\Pr$. This is consistent
with the value of the filling factor of the downflow, $f_d$ 
which does not change in this set of simulations (\Tab{table1}).
The contribution from the thermal conductive flux ($F_{\rm cond}$) becomes increasingly negligible
with the increase of $\Pr$. This is already seen in \Fig{fig:fluxesR0} that $F_{\rm cond}$ is negligible in Run~R0P2.
The disappearance of $F_{\rm cond}$ with the increase of $\Pr$ is really a consequence 
of the increasing $\Ra$.  As $\Ra$ is increased, $F_{\rm enth}$ dominates over $F_{\rm cond}$
which is well-known in solar convection simulations \citep{Miesch05}.
Therefore, in all these simulations (R0P1--R0P20)
we expect, $L_\odot / 4 \pi r^2  \approx {F_{\rm enth}} = v_r T$ to hold.
Here $T$ is the temperature fluctuations relative to the background
which may be also regarded as the temperature deficit in the plumes relative to their surrounding.
Therefore, in these runs, increasing $\Pr$ (i.e., decreasing $\kappa$), increases the temperature fluctuations ($T$)
in plumes. Thus, to transport the same solar luminosity by the enthalpy flux, the convective velocity $v_r$ 
must go down. 

\begin{figure*}
\includegraphics[scale=0.3]{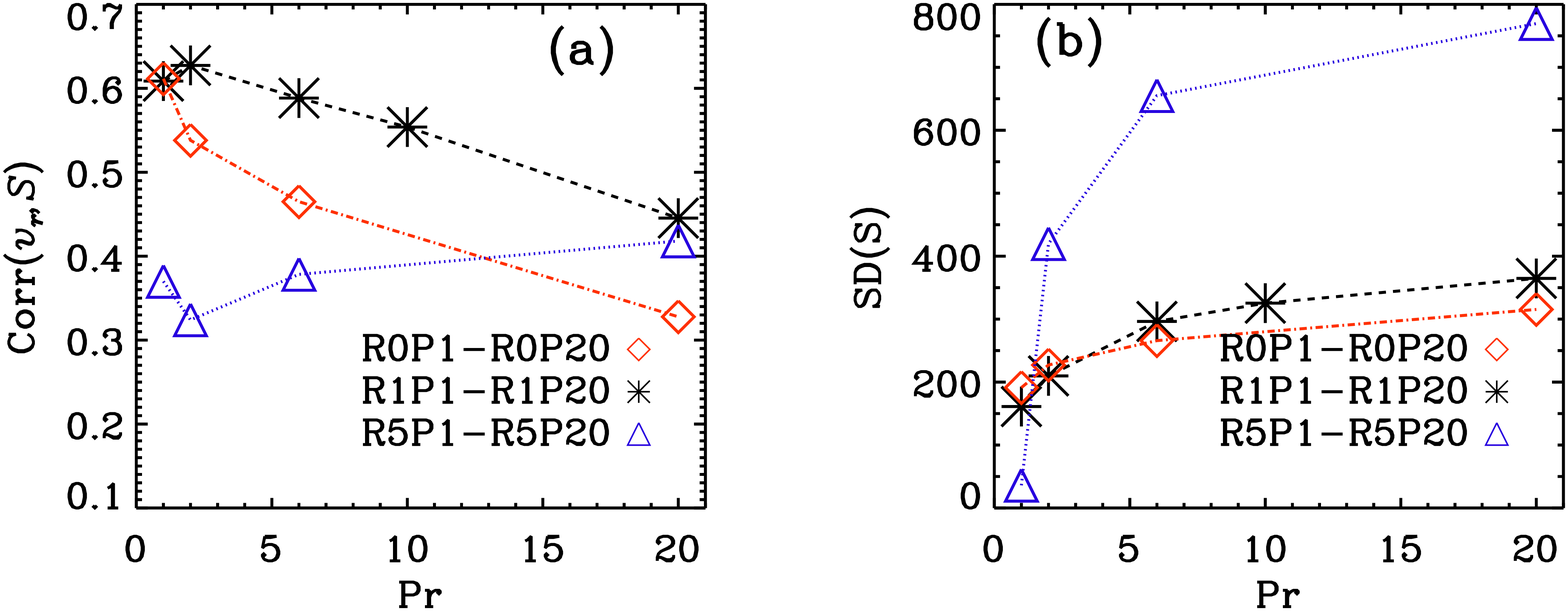}
\caption{
$\Pr$ dependences of (a) linear Pearson's correlation coefficient between 
the radial velocity $v_r(r,\theta,\phi)$ and the entropy fluctuation $S(r,\theta,\phi)$, i.e.,
$\brac{ (v_r - \brac{v_r}_{\theta,\phi}) (S - \brac{S}_{\theta,\phi}) }_{\theta,\phi} / [\mathrm{SD}(v_r) \mathrm{SD}(S)] $, where 
SD($x$) is the  standard deviations of $x$ and thus 
SD($S$) = $\sqrt{ \brac{ [ S - \brac{S}_{\theta,\phi} ]^2 }_{\theta,\phi} }$,
(b) SD($S$) in erg~g$^{-1}$K$^{-1}$.
Both Corr($v_rS$) and SD($S$) are obtained by taking averages 
of values computed at 
$r/R_\odot = 0.712$, 0.794, 0.852, 0.909, 0.989, and 0.997.
}
\label{fig:corr}
\end{figure*}

In computing the correlation between radial velocity and entropy fluctuations,
we find that it slowly decreases with the increase of $\Pr$; see \Fig{fig:corr}(a).
However, the entropy fluctuation, as measured by the standard deviation of $S$
over the horizontal surfaces, 
expectedly increases with the increase of $\Pr$; see \Fig{fig:corr}(b).
This confirms that the decrease of convective velocity is not caused by the
increase of correlation between entropy fluctuation and velocity, rather 
it is due to the increased temperature fluctuations as discussed above.

The behavior is different in rotating simulations because rotation
suppresses the convective motion as well as
inhibits the downward motion of the low entropy plumes.
Therefore in these runs, $F_{\rm cond}$ is {\it not} negligible.
It transports a significant fraction of the total flux as seen in \Fig{fig:fluxes}(a) 
for Runs R1P1 and R1P2.
As the rotation tends to decrease the up/down flow asymmetry by limiting the horizontal size of convection cells \citep{Brown08}, $F_{\rm KE}$ is significantly reduced in rotating simulations. The filling factor is accordingly increased (meaning more symmetric up/down flows) in rotating simulations; see \Tab{table1}.
Hence, for $\Pr <6$ in the R1 set of simulations, we have,
the total flux = $L_\odot / 4 \pi r^2$  $\sim F_{\rm enth} + F_{\rm cond}$.
(Again we ignore $F_{\rm heat}$ here.).
When $\Pr$ is increased from 1 to 2 i.e., from R1P1 to R1P2,
$F_{\rm cond}$ is accordingly suppressed. This causes $F_{\rm enth}$ to
increase, resulting in a higher convective flow speed.

\begin{figure}
\includegraphics[scale=0.4]{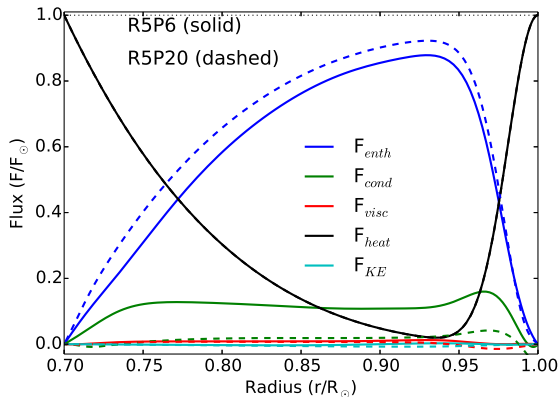}
\caption{Similar to \Fig{fig:fluxes}(b) but obtained from Runs~R5P6 (solid lines) and R5P20 (dashed).
}
\label{fig:fluxesR5}
\end{figure}

However, when $\Pr$ exceeds about 6 in the R1 set of simulations,
 the decrease of convective amplitude can be understood based on the previous
discussion presented for the non-rotating case. This is because in this regime (R1P6--R1P20),
$F_{\rm cond}$ becomes negligible and the relation $L_\odot / 4 \pi r^2  \approx {F_{\rm enth}} = v_r T$
tends to hold in the mid CZ.
This is already seen in \Fig{fig:fluxes}(b); also see values of
$L\rm{_e^p}$ in \Tab{table1}.
Therefore, in all these runs, increasing $\Pr$ further decreases
the convective velocity by increasing the temperature fluctuations $T$. The rate of velocity suppression is smaller in the R1 set of simulations
in comparison to the non-rotating counterpart because, for the same Pr, ${F_{\rm cond}}$ is always higher in the rotating case.  This is better reflected in the rapidly-rotation simulations (R5P1--R5P20).
In \Fig{fig:fluxesR5} we observe that even at $\Pr$ 6, $F_{\rm cond}$ is not negligible
as was the case for Run~R1P6 (\Fig{fig:fluxes}(b)).
Thus when $\Pr$ is increased to 20 from 6 in the R5 set of simulations, the reduction of $F_{\rm cond}$
increases  $F_{\rm enth}$.  This only reduces the convective velocity amplitude slightly (\Fig{fig:urms}).

\begin{figure}
\includegraphics[scale=0.5]{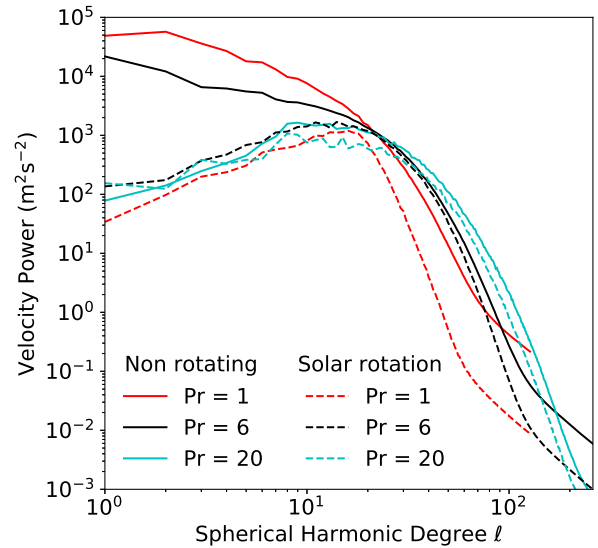}
\caption{Convective power spectra computed at $r = 0.74 R_\odot$ from non-rotating (Runs~R0P1, R0P6 and R0P20; 
solid lines)
and rotating (Runs~R1P1, R1P6 and R1P20; dashed lines) simulations.
Each curve is obtained by averaging many spectra over multiple convective turnover 
timescales $d/\urms$ (18, 67, 18, 4, 16, and 2 for R0P1, R0P6, R0P20, R1P1, R1P6, 
and R1P20, respectively).
}
\label{fig:spectrum}
\end{figure}

 \Fig{fig:spectrum} shows the convective power spectra for the non-rotating (R0)
and rotating (R1) sets of simulations. 
With the increase of $\Pr$, 
convective power at smaller scales (larger horizontal wave numbers) increases.  
This is another indication of the increasing influence of plumes over banana cells as seen previously in \Fig{fig:Spert}.
In other words, the effective size of turbulent eddies decreases with 
the increase of $\Pr$\citep{Omara}.

In non-rotating simulations, at larger-scales (smaller-wave numbers), the convective power decreases rapidly with the increase of $\Pr$.
In the rotating simulations, the large-scale power is relatively low for all simulations.
This is very encouraging because it helps in solving the convective conundrum
which says that the observed solar convection is small at large scales \citep{Hanasoge12,Lord,Greer}. 
The Sun is rotating and thus if we believe that the effective $\Pr$ of the solar convection 
is large, then the convective conundrum might be solved.
We note that this result of large-scale power suppression was also explored by 
\citet{FH16}, who found that the peak of the power spectrum 
shifts toward smaller scales with the increase of rotation rate.

In R1 set of simulations, the large-scale power shows some variation.
The power increases first from R1P1 to R1P6 and then it tends to decrease again (\Fig{fig:spectrum}).
Furthermore, the power increases at small scales (i.e, the increasing influence of plumes over banana cells; \Fig{fig:Spert}) 
with the increase of $\Pr$.
This might change the convective Reynolds stress which is responsible for transporting the angular momentum.
Therefore any change in the convective power may also lead to a change in the differential
rotation pattern that is developed in these rotating convection simulations. 
This we explore in the next section.

\begin{figure*}
\includegraphics[scale=0.45]{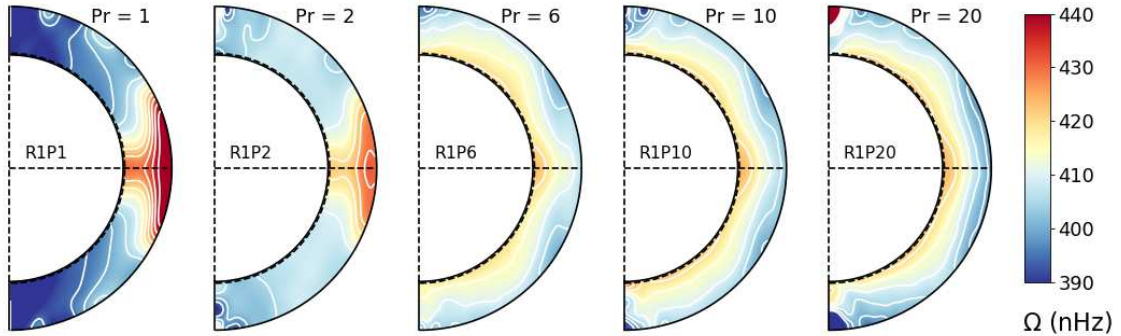}
\caption{
Temporally-averaged rotation rate $\Omega(r,\theta)$ in nHz 
from Runs~R1P1--R1P20 (left to right).
}
\label{fig:DR}
\end{figure*}

\subsection{Differential Rotation}
\label{sec:DR}
\Fig{fig:DR} displays the temporally-averaged differential rotation $\Omega(r,\theta) = \brac{u_\phi}_{\phi,t}/(r \sin\theta)$.
We find that for $\Pr=1$ and 2, the differential rotation is solar-like in the sense that the equatorial region
rotates faster than the higher latitudes and $\Omega(r,\theta)$ increases radially outward.
However, for $\Pr\ge6$ the rotation profile changes to anti-solar because $\Omega(r,\theta)$ is smaller in the low latitudes compared to higher latitudes; 
see values of $\Delta_\Omega^{\mathrm (r)}$ and $\Delta_\Omega^{\mathrm (\theta)}$ in \Tab{table1},
characterizing the features of differential rotation.
It is interesting to note that the $\Ro$ computed based on $\urms$ has not changed
from the simulation at $\Pr = 2$ to 6 (Runs~R1P2 and R1P6 in \Tab{table1}) but the rotation profile has flipped from solar to anti-solar. 
Furthermore, we notice that from $\Pr =6$ (Run~R1P6) to 20 (Run~R1P20), 
$\Ro$ decreases and thus the convection ostensibly becomes more rotationally constrained.
Therefore, we expect an increasing tendency for solar-like differential rotation as in previous global convection simulations \citep{Brown08,gastin,FM15}. However, the result is opposite. The most dramatic demonstration of this is seen when comparing runs for $\Pr = 1$ (R1P1) and $20$ (R1P20).
In these two runs, $\urms$ and $\Ro$ are the same 
but the rotation profiles are completely opposite. 

\begin{figure}
\includegraphics[scale=0.35]{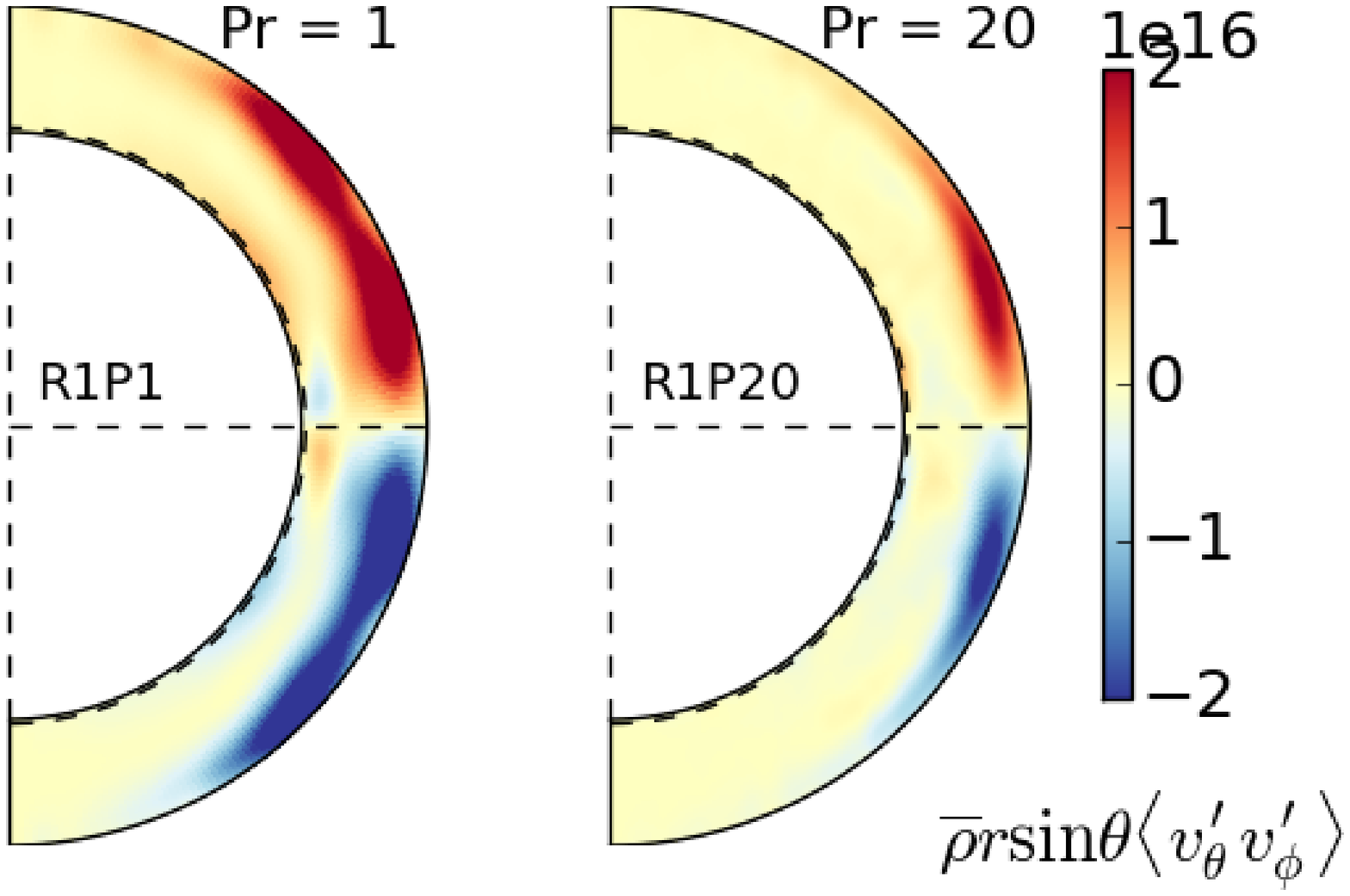}
\caption{
The latitudinal Reynolds stress $\overline{\rho} r \brac{v_\theta^{\prime} v_\phi^{\prime}}_{\phi,t}$ 
(in g~s$^{-2}$)
from Runs~R1P1 (left) and R1P20 (right).
}
\label{fig:tRS}
\end{figure}

Some previous studies have shown that within a certain parameter regime, the rotating convection 
supports hysteresis \citep{gastin,Kap14,Kar15}. That is, if the simulation is 
started from an anti-solar differential rotation, then it produces anti-solar differential
rotation and otherwise a solar-like differential rotation is produced if started from an uniform rotation. 
However, this is not happening here
because all simulations are started from the same initial state of uniform rotation.

\begin{figure*}
\includegraphics[scale=0.35]{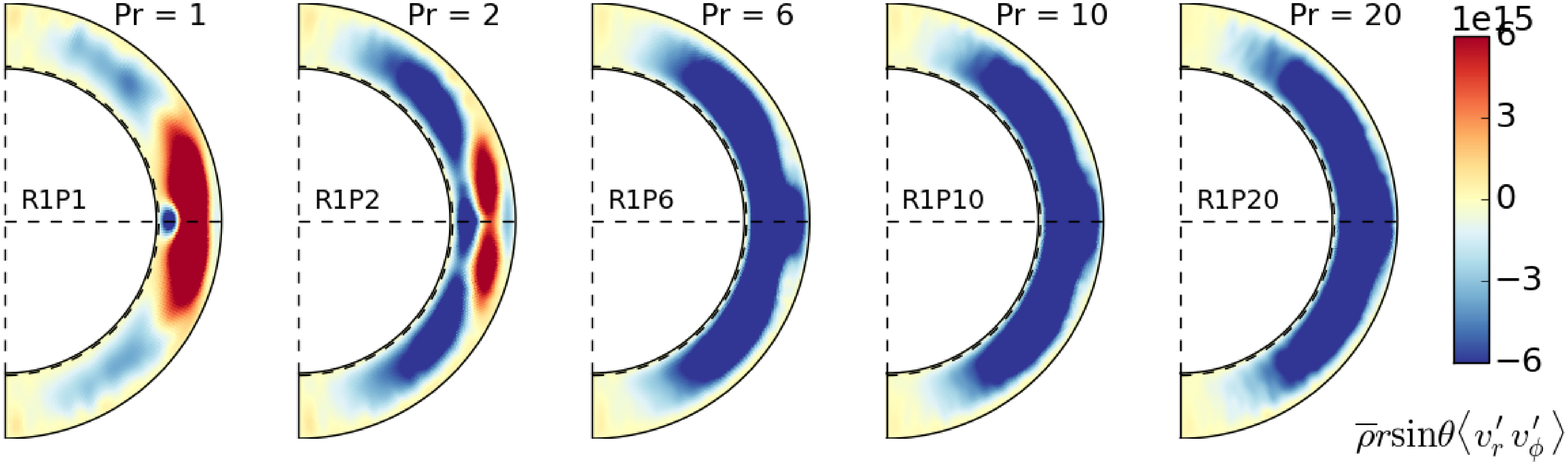}
\caption{
The radial Reynolds stress $\overline{\rho} r \brac{v_r^{\prime} v_\phi^{\prime}}_{\phi,t}$ 
(in g~s$^{-2}$)
from Runs~R1P1--R1P20 (left to right).
}
\label{fig:rRS}
\end{figure*}

The change in the differential rotation profile with the increase of $\Pr$ is caused by the inward angular momentum transport by the downward plumes. As we have discussed above, with
the decrease of the heat diffusivity, the downward motion of plumes becomes stronger; see the increase of $\Pe$ with the increase of $\Pr$ in \Tab{table1}.
Since plumes tend to conserve angular momentum locally (in contrast to banana cells), 
this produces inward angular momentum transport ($\brac{v^\prime_r v^\prime_\phi} < 0$).  
This inward angular momentum transport in turn is responsible for establishing the anti-solar differential rotation \citep{gastin,FM15}.

To confirm this, we compute the radial and latitudinal components of the Reynolds stress, 
$\overline{\rho} r \sin\theta \brac{v_r^{\prime} v_\phi^{\prime}}_{\phi,t}$, 
and $\overline{\rho} r \sin\theta \brac{v_\theta^{\prime} v_\phi^{\prime}}_{\phi,t}$, respectively.
We find that in all simulations, the latitudinal component of Reynolds stress is positive, meaning an equatorward transport of angular momentum; \Fig{fig:tRS}.
However, the value decreases with the increase of $\Pr$; see $\overline{\rm RS}_\theta$ in \Tab{table1}. 
The radial Reynolds stress $\overline{\rho} r \sin\theta \brac{v_r^{\prime} v_\phi^{\prime}}_{\phi,t}$
shows a different behavior. Only for Runs R1P1 and R1P2, it is positive while for all other runs it is negative (\Fig{fig:rRS}). 
In fact, already for 
Run~R1P2, $\overline{\rho} r \sin\theta \brac{v_r^{\prime} v_\phi^{\prime}}_{\phi,t}$ 
it is negative in most of the CZ, except in low latitudes.
It is the inward angular momentum transport due to plumes that makes the differential rotation anti-solar.

As mentioned above, $\Ro$ is ostensibly smaller for the simulations at higher $\Pr$.  
However, these simulations exhibit more power at large wavenumbers, 
where the effective $\Ro$ is large (\Fig{fig:Spert}).  
This excess power at small scales is another indication of the increasing 
importance of plumes relative to banana cells as $\Pr$ is increased.

There is a long history of studying the solar to anti-solar differential rotation transition, 
starting from \citet{Gi77}
and more recently by many others \citep{gastin,Kap14,FF14,Kar15,FM15}. These studies have shown that a transition from  solar to 
anti-solar rotation can be achieved only by reducing the rotational effect of the convection
i.e., by increasing $\Ro$. In the past, this has been done either by decreasing the rotation rate\citep[e.g.,][]{Gue13} or by increasing the 
convective velocity \citep[e.g.,][]{FF14,FM15,Kap14,Kar15}. In our simulations, we find the solar to anti-solar transition
not by increasing the global $\Ro$, but by decreasing the thermal heat diffusivity which increases the inward angular momentum transport
and decreases the equatorward angular momentum transport.

\begin{figure}
\includegraphics[scale=0.6]{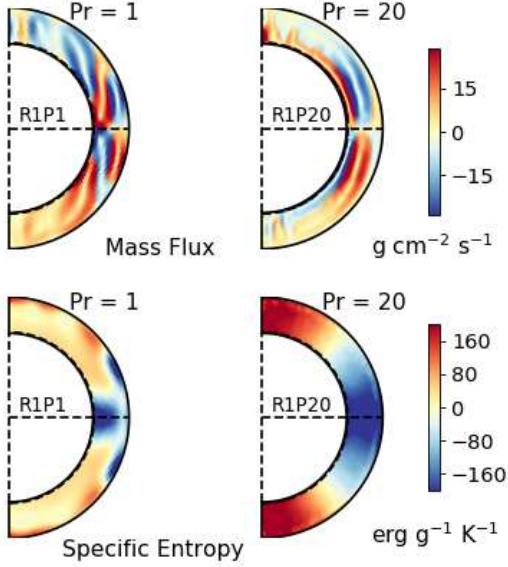}
\caption{
Top: the mass flux ($\rho \sqrt{v_r^2 + v_\theta^2}$)
of meridional flow; blue and red correspond to anti-clockwise and clockwise
meridional circulation.
Bottom: specific entropy $\brac{S}_{\phi,t}$.
Left and right panels are obtained from Runs~R1P1 and R1P20, respectively.
}
\label{fig:MC}
\end{figure}

It is known that convection simulations tend to produce single-cell meridional flow 
when differential rotation is antisolar and multicellular otherwise \citep{Kar15,FM15,Gue13}.
In fact, in some studies, the antisolar differential rotation is attributed to poleward angular momentum transport by the meridional flow \citep{Dob06,FM15}.
Particularly, \citet[][hereafter FM15]{FM15} showed that, in the large-$\Ro$ regime, inward angular momentum transport by the convective Reynolds stress induces a single-cell meridional circulation profile (one cell per hemisphere) through gyroscopic pumping and that this induced circulation is primarily responsible for spinning up the poles relative to the equator.  A similar process is also operating in our high-$\Pr$ simulations but the resulting anti-solar differential rotation profile is very different.  Here inward angular-momentum transport by the increasingly plume-dominated convection induces a single-cell meridional circulation profile much as in FM15 (\Fig{fig:MC}, top row).  However, in FM15, this induced circulation is strong enough to establish a cylindrical, Taylor-Proudman rotation profile with an $\Omega$ toward the rotation axis.  In the simulations reported here, this single-cell circulation (counter-clockwise in the northern hemisphere, clockwise in the southern hemisphere) is suppressed by a strong poleward entropy gradient (\Fig{fig:MC}, lower right) that exerts an opposing baroclinic torque.   The resulting thermal wind balance sustains an $\Omega$ gradient that is approximately radially inward (Fig.\ \ref{fig:DR}, right).  By contrast, the entropy gradient in FM15 is equatorward and too weak to sustain significant departures from a Taylor-Proudman (cylindrical) state.

We attribute this difference to the rotational influence and the subadiabatic stratification.  In FM15, as here, downward plumes are responsible for the inward angular momentum transport.  However, in the simulations reported here, the transition to plume-dominated convection is achieved by an increase of the Prandtl number as opposed to an increase in the Rossby number.   So, the rotational influence remains strong enough to make the convective heat flux at the poles (where plumes are not deflected by the Coriolis force) more efficient than at lower latitudes \citep{Kit95,Miesch00}.  This contributes to the poleward entropy gradient, particularly in the upper convection zone.  In the lower convection zone, the subadiabatic stratification established by our high-$\Pr$ simulations prevents the positive feedback between mechanical and thermal driving that can otherwise amplify the meridional circulation.  This can be illustrated by first considering the opposite case of a superadiabatic convection zone.  Here the circulation pattern established by gyroscopic pumping carries high-entropy fluid upward at the equator and low-entropy fluid downward at the poles.  This warms the equator relative to the poles and thus acts to buoyantly enhance the circulation.  In other words, gyroscopic pumping can trigger an axisymmetric convective instability that proceeds until viscous and thermal dissipation eventually halt its growth (FM15).  By contrast, subadiatatic stratification suppresses this axisymmetric convection mode.  Upward advection of low-entropy fluid at the equator and downward advection of high-entropy fluid at the poles establishes a poleward entropy gradient that baroclinically opposes the induced circulation and establishes thermal wind balance.  In Rempel's\citep{rempel05} mean-field models, this process was put forth as a promising mechanism for establishing a solar-like differential rotation profile with an equatorward (non-cylindrical) $\Omega$ gradient. Our simulations suggest that a high effective Prandtl number may support this picture by producing the requisite sub-adiabatic stratification.  However, a solar-like rotation profile will only be established if the convective angular momentum flux has a sufficiently strong equatorward component in addition to an inward component.

\begin{figure}
\includegraphics[scale=0.6]{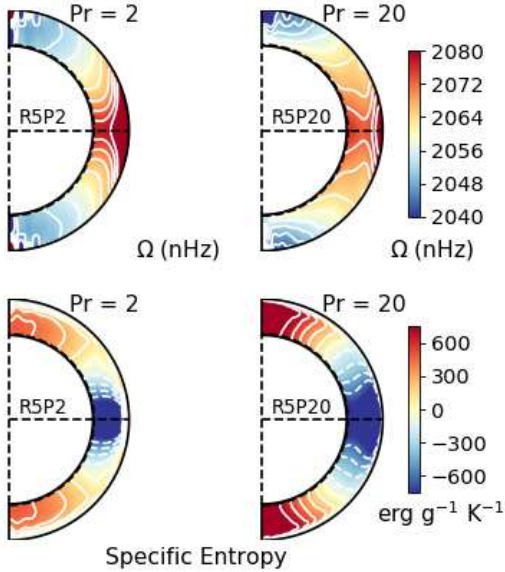}
\caption{
Top: angular frequency $\Omega$, bottom: 
specific entropy $\brac{S}_{\phi,t}$.
Left and right panels are obtained from Runs~R5P2 and R5P20, respectively.
}
\label{fig:nonTP}
\end{figure}

In rapidly rotating simulations (Runs~R5P1--R5P20),
the convective velocity is significantly suppressed 
(in addition to the higher $\Omega$)
and thus the
convection is more rotationally constrained (having much smaller $\Ro$).
Hence, a solar-like differential rotation is maintained in all Runs~R5P1--R5P20; see \Tab{table1}
and top panels of \Fig{fig:nonTP}.
Again, from \Fig{fig:nonTP}, we confirm that the latitudinal entropy difference 
is enhanced as $\Pr$ is increased, and as a result, a conical profile 
of differential rotation is obtained in Run~R5P20.

\section{Summary and Conclusions}

We have considered the recently highlighted problem
that solar convection simulations might be overestimating the convective power
at large scales relative to observations \citep[e.g.,][]{Hanasoge12,Lord,Greer}. 
Recent studies have suggested that due to strong magnetic field
produced by the small- and large-scale dynamos, the solar convection might be operating at large effective
$\Pr$  and this could suppress the convective velocity amplitude \citep{HRY15,Omara,Bekki}.
In this paper we address this problem for the first time using global convection simulations under the influence of rotation. 
 We explore the effect of rotation on the convective amplitude and the differential rotation 
and we demonstrate that what is beneficial for one may not be beneficial for the other. 
 In particular, though a high effective $\Pr$ can reduce the amplitude of large-scale convective motions, 
it can also give rise to an anti-solar differential rotation.

As in previous simulations \citep{Omara,Bekki}, we consider non-magnetic (hydrodynamic) 
simulations with $\Pr > 1$, which is intended to mimic the influence of small-scale magnetism.
In our simulations, the convection is excited by introducing a heating term in the entropy equation
(which mimics the radiative energy flux tabulated in the standard solar model; see \citet{FH16,Omara}) and a cooling term
(which captures the efficient surface cooling near the surface; see \citet{HRY14,Bekki}). We conduct a series of global
hydrodynamic simulations at varying $\kappa$ i.e., at varying $\Pr$ and at varying rotation rate.

As reported in previous local convection simulations by \citet{Bekki}, a subadiabatic layer (with $\delta < 0$)
in the lower CZ is formed in our non-rotating simulations (R0P1--R0P20)
due to continuous deposition of the low entropy downward plumes. The depth of the
subadiabatic layer is about 90~Mm for Runs~R0P1--R0P2 and it is bigger at larger $\Pr$.
However, in simulations at the solar rotation rate, the subadiabatic layer is formed only for $\Pr>2$.
The reason is that the rotation inhibits the inward transport of the cool downward plumes and
thus tries to inhibit the formation of subadiabatic layer. This effect is stronger
in the more rapidly-rotating simulations and therefore we observe a thin subadiabatic layer only in Runs~R5P6 and R5P20.

In the non-rotating set of simulations (R0P1--R0P20), the convective velocity systematically decreases with the increase of $\Pr$,
in agreement with previous studies \citep{Omara,Bekki}. However, in rotating sets (R1, and R5), we find a different behavior.
At lower $\Pr$, the convective velocity increases with the increase of $\Pr$ and then above a certain $\Pr$, depending on the rotation rate,
it decreases but at a slower rate than the non-rotating set. In the lower $\Pr$ range, the thermal conduction carries a significant
amount of the solar luminosity
and this contribution decreases with the decrease of $\kappa$ (i.e., increase of $\Pr$). This decrease of $F_{\rm cond}$ with the increase of $\Pr$, causes the enthalpy flux ($F_{\rm enth}$) to increase and thus also the convective velocity.

When $\kappa$ is sufficiently small,
the increase of $\Pr$ does not increase $F_{\rm enth}$ rather it increases
thermal fluctuations in the cool downflow plumes. This 
allows the convection to carry the solar luminosity at smaller
speed.
In the solar CZ, $\kappa$ is very small and thus the low-$\kappa$ regime is the one 
that is most relevant to the Sun. In this regime, the increase of $\Pr$ causes to decrease the convective flow speed.

With the increase of $\Pr$ the convective power increases at small scales and 
the behavior is different at large scales.
In all rotating simulations, independent of the value of $\Pr$, the convective power is smaller by about two orders of magnitude than that in non-rotating simulations at low $\Pr$.
 
The most interesting result of our rotating simulations is the behavior of differential rotation
with the increase of $\Pr$. At large $\Pr$ (small $\kappa$), cool downward plumes that are formed near the surface
can reach all the way to the lower CZ without losing their thermal content. This downward propagation of low entropy plumes
transport angular momentum radially inward ($\brac{v_r^{\prime} v_\phi^{\prime}} < 0$), 
offsetting the equatorward angular momentum transport by banana cells.
This inward angular momentum transport by the plumes promotes
an anti-solar differential rotation. 

In previous work, it had been assumed that the decrease in convective velocity amplitude with increasing Pr would help promote solar-like differential rotation by increasing the rotational influence (lower Ro).  However, we find that the lower amplitude of convective motions is accompanied by a change in the convection structure that is increasingly influenced by small-scale plumes.  And, such plumes tend to transport angular momentum inward, establishing a differential rotation profile in striking contrast to the solar rotation profile. Such minimally diffusive (low $\kappa$) plumes are likely also present in the solar convection zone, driven by radiative cooling in the photospheric boundary layer.  So, our results cast doubt on the idea that a high effective Prandtl number may be a viable solution to the solar convection conundrum.  Furthermore, our work emphasizes that any resolution of the convection conundrum that may be proposed in the future must take into account angular momentum transport as well as heat transport.  This is particularly true for models that rely on small-scale convective plumes.

{\bf ACKNOWLEDGMENTS}\\
We thank the referees for carefully reading the manuscript 
and providing valuable comments which help to improve the clarity of the presentation. 
B.B.K. was supported by the NASA Living With a Star Jack Eddy Postdoctoral 
Fellowship Program, 
administered by the University Corporation for Atmospheric Research. 
The code Rayleigh used in this study
has been developed by Nicholas Featherstone with support by the National Science
Foundation through the Computational Infrastructure for Geodynamics (CIG).  This
effort was supported by NSF grants NSF-0949446 and NSF-1550901.
Y.B.'s visit to HAO was supported by the 
Leading Graduate Course for Frontiers of Mathematical Sciences and Physics of the University of Tokyo.
The National Center for Atmospheric Research is 
sponsored by the National Science Foundation. 
Computations were carried out with resources provided by NASA’s 
High-End Computing program (Pleiades).

\bibliography{paper}

\end{document}